\documentstyle[11pt,aaspp4]{article}
\nonstopmode	% don't stop on errors, useful for running `latex' from `make'

\def\bibcode#1{}
\def\etal{{et al.\hskip 3pt}}
\def\eg{{\it e.g.\hskip 2pt}}
\def\ie{{\it i.e.\hskip 2pt}}
\def\kmsMpc{{\rm km}\,{\rm s}^{-1}\,{\rm Mpc}^{-1}}
\def\Junits{{\rm erg\,s}^{-1}\,{\rm cm}^{-2}\,{\rm sr}^{-1}\,
                  {\rm Hz}^{-1}}

\newcommand{\be}{\begin{equation}}
\newcommand{\ee}{\end{equation}}

\singlespace
\slugcomment{Submitted to New Astronomy}

\lefthead{Dav\'e, Dubinski \& Hernquist}
\righthead{Parallel TreeSPH}

\begin{document}

\title{Parallel TreeSPH}

\author{Romeel Dav\'e, John Dubinski\altaffilmark{1}, and Lars Hernquist\altaffilmark{2}}
\affil{Astronomy Dept., University of California, Santa Cruz, CA 95064}
\altaffiltext{1}{Current address: CITA, Univ. of Toronto, 60 St. George St., Toronto, Ontario M55 IA7, Canada}
\altaffiltext{2}{Presidential Faculty Fellow}
%%%%%%%%%%%%%%%%%%%%%%%%%%%%%%%%%%%%%%%%%%%%%%%%%%%%%%%%%%%%%%%%%%%%%%%%

%%%%%%%%%%%%%%%%%%%%%%%%%%%%%%%%%%%%%%%%%%%%%%%%%%%%%%%%%%%%%%%%%%%%%%%%
\abstract
We describe PTreeSPH, a gravity treecode combined with an SPH hydrodynamics code
designed for massively parallel supercomputers having distributed memory.
Our computational algorithm is based on the 
popular TreeSPH code of Hernquist \& Katz (1989).
PTreeSPH utilizes a domain decomposition procedure and a synchronous
hypercube communication paradigm to build self-contained subvolumes
of the simulation on each processor at every timestep.  Computations then 
proceed in a manner analogous to a serial code.  We use the Message Passing
Interface (MPI) communications package, making our code easily
portable to a variety of parallel systems.
PTreeSPH uses individual smoothing lengths and timesteps, with a communication
algorithm designed to minimize exchange of information while still
providing all information required to accurately perform SPH computations.
We have additionally incorporated cosmology,
periodic boundary conditions with forces calculated
using a quadrupole Ewald summation method, and radiative cooling
and heating from a parameterized ionizing background
following Katz, Weinberg \& Hernquist (1996).
The addition of other physical processes, such as star formation, is straightforward.
A cosmological simulation from $z=49$ to $z=2$ with $64^3$ gas particles 
and $64^3$ dark matter particles requires $\sim 6000$ node-hours
on a Cray T3D, with a communications overhead of $\sim 10$\% and is
load balanced to a $\sim 90$\% level. 
When used on the new Cray T3E, this code will be capable of performing cosmological
hydrodynamical simulations down to $z=0$ with
$\sim 2\times 10^6$ particles, or to $z=2$ with $\sim 10^7$ particles,
in a reasonable amount of time.  Even larger simulations will be practical
in situations where the matter is not highly clustered or when periodic
boundaries are not required.

\bigskip
\bigskip
\noindent {\it Subject Headings:} methods: numerical --- cosmology: theory

\newpage

\section{Introduction}

Gas dynamics and gravitational forces together govern 
the evolution of astrophysical systems on nearly all scales.
Star and planet formation are thought to occur in accretion disks well-represented
as collapsing gas clouds in quasi-hydrostatic equilibrium.
The formation and evolution of star clusters is governed by 
interactions with the interstellar medium, as well as viscous
forces generated from tidal perturbations in individual stellar interactions.
Shock heating from supernovae feedback is important for the observed structure 
of galaxies and the creation of hot X-ray halos around clusters.
All the galaxies we see are themselves thought to have resulted from
dissipational collapse through heirarchical gravitational instability.
On these largest scales, comparison between observation
and theory requires some understanding of the interplay between
dissipationless dark matter and baryonic material.

Analytical treatments of the processes relevant to cosmology are generally
applicable only when the matter is still evolving in a
linear or quasi-linear fashion, or more generally for objects possessing 
inherent symmetries such as stars, supernovae, or accretion disks.
However, many systems are intrinsically asymmetric and/or highly nonlinear,
such as galaxies forming from primordial fluctuations.
Semi-analytical treatments of galaxy formation 
invariably make {\it ad hoc} assumptions about the relationships between
gas and dark matter in the highly nonlinear regime 
(\eg \cite{kau93}; \cite{hey95}; \cite{som97}).
A proper understanding of galaxy and structure formation requires that
dissipation, shocks, and pressure forces be taken into account,
in addition to gravitational collapse.

Highly nonlinear systems are best modelled numerically by evolving
large numbers of particles and/or grid cells
self-consistently under both gravitational and
hydrodynamical forces.  A variety of techniques to do this have been developed,
each with its own advantages and disadvantages.  Gravitational forces are
computed using either grid-based or particle-based algorithms.
The simplest particle-based technique directly sums pairwise
forces between all particle pairs.  These codes are
useful for many purposes (\eg \cite{aar85}), but the computational time grows
with particle number $N$ as $O(N^2)$ or worse, 
making them appropriate only for small systems ($N\la 10^4$).
Tree codes place particles in a hierarchical data structure, 
and use multipole expansions to approximate the force between
distant groups of particles, reducing the scaling to $O(N\log{N})$ (\eg 
Barnes \& Hut 1986, hereafter \cite{bar86}).
In particle-mesh (PM) codes, the gravitational potential is
computed on a grid using Fast Fourier Transforms (FFTs), which
also scales as $O(N\log{N})$ but with fewer operations than a treecode
(\eg \cite{hoc80}).
The main drawback of this approach is that the resolution is limited by the cell size.  
Adaptive mesh 
codes have been developed to subdivide or deform cells in dense regions
to obtain better dynamic range (\cite{vil89}; \cite{pen95}; \cite{xu97}).
Also, hybrid codes such as PP-PM (P$^3$M; \cite{efs85}) and Tree-PM 
(TPM; \cite{xu95}) alleviate resolution limitations.  Other techniques,
such as the self-consistent field (SCF) method (\eg \cite{her92}) can
be even faster, but are generally useful only for systems not far
from a well-specified equilibrium.

Hydrodynamical forces can be computed in either a Lagrangian or Eulerian manner.
Eulerian codes represent the fluid on a grid of cells, 
and compute the flux of fluid across cell boundaries, as in
the Piecewise Parabolic Method (PPM) (\cite{col84}; \cite{bry94}).
In Lagrangian codes, the dynamical equations are obtained from the 
Lagrangian form of the hydrodynamical conservation laws.
Some Lagrangian codes represent
the fluid by particles without the use of a grid, as in
Smoothed Particle Hydrodynamics (SPH) (\cite{luc77}; \cite{gin77}).

In principle, any gravity solver may be combined
with any hydrodynamics method.  Grid-based codes include PM-PPM
(\cite{bry94}), PM-TVD (\cite{ryu93}), and 
adaptive mesh hydrodynamics codes (\eg \cite{bry95}).
Lagrangian examples include P$^3$M-SPH (\cite{evr88}; \cite{cou95}),
and GRAPE-SPH (\cite{ste96}), which uses the special-purpose GRAPE
hardware to perform rapid pairwise gravity summation.
Lagrangian codes provide much better spatial resolution in high
density regions compared with Eulerian codes for a given
computational expense, at the cost of poorer shock resolution
and lower resolution in underdense regions (\cite{kan94}).
For many astrophysical applications the overdense regions are of 
most immediate interest, and in those cases Lagrangian codes are
preferable if shocks do not dominate the dynamics of the system.

In this paper we focus on a pure particle-based combination of
gravity and hydrodynamics solvers
analogous to the TreeSPH code of Hernquist \& Katz (1989; hereafter \cite{her89}).
In SPH, the gas is sampled and represented by particles, which
are smoothed to obtain a continuous distribution of gas properties.
Since there is no grid, there are no inherent constraints on the
global geometry or spatial resolution.  
Neighbor finding is also done using a tree structure, and thus the
entire code scales as $\sim O(N\log N)$.  
However, unlike grid-based codes, SPH cannot handle
arbitrarily large gradients due to its finite particle resolution.
Also, an artificial viscosity is used
to capture shocks, further limiting the spatial
resolution locally.  Despite these compromises, TreeSPH has been successfully
used in a wide range of astrophysical applications, including
giant molecular clouds (\cite{gam92}), 
colliding galaxies (\cite{mih94}), 
ram pressure stripping in clusters (\cite{kun93}),
formation of galaxies (\cite{kat91}) and cluster (\cite{kat93}),
and the high-redshift Lyman alpha forest (\cite{her96}).

A major goal in numerical astrophysics is to improve the dynamic range
of simulations.  One would ideally like to simulate volumes comparable to the size of
the Universe ($\sim 10^3$ Mpc), but resolve star forming regions
in galaxies on the parsec scale.  The spatial dynamical range required 
per dimension is thus $\sim 10^9$, well beyond the
$\sim 10^3$ in dynamic range that codes can currently achieve.
High resolution studies of even single collapsing protogalaxies
in a cosmological setting
require a dynamic range $\ga 10^4$.
TreeSPH endows each gas particle with its own smoothing
length and timestep, thus
improving the dynamic range substantially by being adaptive in both
space and time.  TreeSPH has been vectorized efficiently 
(\cite{her89}; \cite{her90}), but
even on vector supercomputers the largest TreeSPH cosmological 
simulation to date 
(Katz, Weinberg \& Hernquist 1996, hereafter \cite{kat96}) 
employed a total of $\approx 524,288$ particles with a dynamic range of $\sim 2000$.

Massively parallel supercomputers (MPPs) link hundreds of workstation
processors together to yield an overall computational power
more than an order of magnitude greater than that of current vector supercomputers.
While MPPs are attractive, there are a number of major difficulties in adapting codes
to run on these machines, often requiring significant algorithmic changes 
from serial or vector code.  The first difficulty is that in 
distributed memory systems, each processor 
possesses only a relatively small amount of local memory, and accessing
information from another processor's memory is slow compared to
the computation speed.  Thus a parallel code must 
subdivide a simulation and exchange information between processors in a
manner which minimizes communication time,
while not taxing each processor's memory.
The second complication is {\it load balancing}, \ie insuring that
no processor spends a significant amount of time idly waiting
for another processor to send required information.  Good load balancing 
can be achieved either by designing an algorithm so that each processor has
roughly an equal amount of computational work between {\it synchronous} communications, 
or by implementing
an {\it asynchronous} communication scheme by which processors continue
to do other computations while waiting to send or receive information.

In this paper we present a parallel implementation of a gravitational
treecode combined with SPH, called PTreeSPH.  
While the underlying numerical techniques are
similar to those in TreeSPH, our implementation on MPP machines required a 
complete redesign of the code as well as several major algorithmic changes.
PTreeSPH is a $C$ code which uses a domain decomposition prescription to subdivide
the simulation and a synchronous hypercube message-passing paradigm to
build small ``locally essential" simulation subvolumes on each processor.
The $N$-body portion of the code was developed by Dubinski 
(1996; hereafter \cite{dub96}), to which we have added a parallelized version of SPH
as well as the dynamics required for cosmological simulations.
Building locally essential simulations on each processor allows the 
parallelization to be decoupled from the computations, making it
straightforward to incorporate additional physical processes.
In addition, information transfers occur in single bursts rather
than continually during a simulation, thereby lowering communication overhead.
We utilize the Message Passing Interface (MPI) package exclusively to handle 
processor communications, which makes the code 
easily portable to most major parallel supercomputers, including the Cray
T3D/T3E, IBM SP2, and Intel Paragon, shared memory systems such as the
SGI Power Challenge, as well as to networks of workstations.
While our parallel algorithm could likely be made more efficient by tailoring
it to specific machines and including some asynchronous communication,
we are more interested in producing a reasonably efficient code
which is portable and adaptable to a wide variety of applications.

Several other groups are now parallelizing SPH in various forms.  
The Virgo Consortium has recently developed a parallel P$^3$M-SPH code
capable of doing cosmological simulations (\cite{pea95}; \cite{jen96}).
Evrard and Brieu (private communication) are working on a similar code.
\cite{war94} have developed a
generalized parallel $N$-body code using a 
hashed octal tree structure to asynchronously
access information on other processors, and they describe how their
code may be applied to neighbor finding as required in SPH.
Also, \cite{dik97} have written a parallel $N$-body treecode 
similar in spirit to \cite{war94},
and are now adding their own parallel version of SPH to it. 

%%%%%%%%%%%%%%%%%%%%%%%%%%%%%%%%%%%%%%%%%%%%%%%%%%%%%%%%%%%%%%%%%%%%%%%%%%%%%%
\section{The Parallel $N$-Body Treecode}

  \subsection{The Barnes-Hut Treecode}

\cite{dub96} fully describes Dubinski's implementation of a parallel treecode,
which is based on Salmon's (1990) algorithm.
It uses the \cite{bar86} method of tree construction, by which
the simulation volume is subdivided recursively into equal-sized octants,
called {\it cells}.  If there is only a single particle in a given cell, that 
cell is not further subdivided.  The data structure is then a tree,
with the {\it root cell} at the top, and successively smaller cells at
each level.  
The mass distribution within a cell is approximated as
a multipole truncated typically at quadrupole order (\eg \cite{her87}).
The force on any given particle is then the sum of the (exact) force from
nearby particles, plus the force from distant cells approximated as 
multipoles.  The criterion for approximating cells as a
multipole is given by the {\it opening criterion:}
\begin{equation}
d > {l\over\theta} + \delta
\end{equation}
where $d$ is the distance from the particle to the cell's center of mass,
$l$ is the longest edge of the cell, $\delta$ is the distance between
the cell's center of mass and its geometric center, and $\theta$ is 
a user-defined parameter that gives greater accuracy when smaller.  
This opening criterion, introduced by \cite{bar94},
avoids numerical pathologies from highly skewed 
distributions within a cell.  For $\theta \approx 1.0$ at quadrupole order, 
the median error is roughly 0.7\%.
Dubinski also improved efficiency by using grouping and 
a non-recursive treewalk
using a linked list of tree cells (see \cite{dub96} for details).

  \subsection{Domain Decomposition and Locally Essential Trees}

To parallelize a particle simulation, one must have an algorithm
to assign particles to processors.  \cite{dub96} used the intuitive 
{\it domain decomposition} method, whereby a given processor
handles all the particles within a rectangular subvolume of the system.
A treecode naturally lends itself to such a decomposition, since
the processors can be assigned to be cells in an overall tree (see \cite{dub96}: fig.~4).
However, mapping the equal-sized subvolumes inherent in the Barnes-Hut tree 
structure onto individual processors
does not lend itself to optimal load balancing, especially in
clustered simulations typical of gravitational collapse, because
clustered subvolumes require more computational time than unclustered subvolumes.
\cite{sal90} solved this problem by tracking the ``work" done by
each particle (\ie the number of floating point operations required
to compute the force of each particle), and using that information
to subdivide the simulation into volumes of roughly equal computational work (rather than
equal volume) at the next timestep.  This produces an ORB (orthogonal
recursive bisection) load-balanced processor tree.
Initially, all particles are assumed
to require equal work and the simulation may not be well load-balanced, 
but within a few timesteps the load balancing becomes quite satisfactory (\cite{dub96}: Table~1).
On each individual processor, a Barnes-Hut tree is constructed and
appended below ORB tree.

To perform a force calculation, a given processor's particles need
information from trees on other processors.  However, each processor
cannot maintain a full copy of every other processor's tree structure
due to memory constraints.  Instead, each processor builds its own
{\it locally essential tree}, \ie the portion of the entire simulation's
tree which is required for all the local particles' force computations.
The method for doing this is described in \cite{dub96}, and the
reader is referred there for details.  The central idea is that
after all the parallel manipulations and communications, each processor contains
a tree structure which is sufficiently complete to perform all its local particles'
gravitational force computations, and no further communication is required to 
compute the gravitational accelerations for that timestep.

  \subsection{Periodic Boundaries using Ewald Summation}\label{sec: ewald}

\cite{dub97} has recently implemented a cosmological version of the parallel
tree code using the Ewald method
(\cite{ewa21}; Hernquist, Bouchet \& Suto 1991, hereafter \cite{her91}).
Cosmological $N$-body simulations are typically run in a 
cube with periodic boundaries in a comoving coordinate system (\cite{efs85}).
While the periodicity is a natural property of FFT based codes, it must be
imposed in a treecode by replacing the
usual Newtonian gravitational potential 
of a particle (or cell) with the potential of the infinite,
periodic lattice of image particles (or cells).
The potential of the images can be represented accurately
using Ewald's method which replaces the single, slowly converging sum for
the images with two, rapidly converging sums which can be truncated at a
small number of terms.
In practice, the tree descent is modified so that the distance to a cell
used in the cell-opening criterion is the minimum of the distance to the
cell within the simulation cube and the nearest image cell, \ie ${\bf
\delta r}
= [{\rm min}(\delta x,L-\delta x),{\rm min}(\delta y,L - \delta y),{\rm
min}(\delta z, L- \delta z)$], where $L$ is length of the cube.
The potential from a cell is taken as the sum of the Newtonian potential
as calculated by the treecode plus a correction term from the lattice of images.
Since the correction terms are relatively expensive to calculate,
we calculate them  on a grid prior to the simulation
and then use bi-linear interpolation 
to determine the correction terms as needed.
In \cite{her91} and TreeSPH, 
the correction terms are only calculated to monopole order forcing the use
of a small opening-angle parameter.
In PTreeSPH, we calculate the quadrupole order Ewald
correction term as well for particle-cell interactions allowing larger
values of the opening angle parameter and more accurate forces.
It is too costly to use bilinear interpolation for the quadrupole terms but
since they are smaller corrections it is sufficient to use nearest grid
point table lookups for these terms.
The calculation of Ewald corrections slows down the code generally by a
factor of two for a given number of particle-cell interactions.
Typical code speeds on the T3D for the gravity portion are therefore 
between $\sim 100$--200 particles/s/processor depending 
on the value of $\theta$.

The acceleration errors, $\delta a/a$, associated with the Ewald method tend to be
large in the linear regime for a given opening angle parameter $\theta$.
In a 32K particle CDM test simulation, the errors 
for $\theta=0.5(0.7)$
are as large
as 10\%(30\%) at monopole order and 2.5\%(7\%) 
at quadrupole order in the linear regime.  Furthermore, the accelerations
are systematically overestimated, so when $\theta$ is too large 
particles can be
pushed too swiftly through the linear regime with consequences that will
be outlined in \S\ref{sec: cosmo}.
Once the particles become clustered, the errors are much smaller with $\delta/a
= 0.2\%(1.1\%)$ for $\theta=0.7(1.0)$ both to quadrupole order.
We therefore use $\theta = 0.4$ in the linear and weakly
clustered regime, growing with the expansion factor
up to $\theta = 0.8$.
Once the particles are strongly clustered, a value of $\theta=0.8$
gives accelerations with $\sim 0.7$\% accuracy.
Quadrupole order corrections are used at all times.

We have compared the Ewald tree code to two other $N$-body codes: 
TPM (\cite{xu95}) and an adaptive particle mesh (APM) code (\cite{cou95}).
The results of a 2~million particle CDM simulation using TPM and our
code agree very well.  The density profiles of the ten largest virialized
clusters in the simulation are nearly identical.  The only noticeable
differences are in the positions of the small satellites in orbit 
around larger dark halos.  This might be expected since bound orbits
diverge chaotically in the presence of small differences in the
acceleration brought on by random variations.  
A comparison of the formation of a cluster
in a 256K particle simulation with Couchman's code also showed good agreement.  
The cluster density profiles again are nearly identical. 

  \subsection{Performance Characteristics}

Ideally, the wall-clock time required by a parallel code should scale precisely
as the inverse of the number of processors employed.
In practice this perfect scalability cannot be 
achieved because communications and the overhead associated with
locally essential tree building increase with the number
of processors.  Therefore it is desirable to place as large a number
of particles on each processor as memory constraints will allow, so that
computation time versus communication time is maximized.  On a Cray T3D
with 8 Megawords per processor, this translates to over 100K $N$-body particles.
However, even
down to 16K particles per processor Dubinski's treecode barely suffers
a few percent loss in speed from the ideal (\cite{dub96}: fig.~7), showing good
scaling and load balancing.  The total communication overhead is also small, being 
less than 5\% for fully-loaded processors (\cite{dub96}: Table~1; see also
\S\ref{sec: bench} and Table~\ref{table: bench}).

%%%%%%%%%%%%%%%%%%%%%%%%%%%%%%%%%%%%%%%%%%%%%%%%%%%%%%%%%%%%%%%%%%%%%%%%%%%%%%
\section{Parallelizing SPH}

  \subsection{General Overview of SPH}

Smoothed Particle Hydrodynamics (see \cite{mon92a} for a review) uses
locally averaged hydrodynamical quantities to estimate properties associated
with individual particles.  Each particle is smoothed with a
kernel, typically a Gaussian or spline, over a range of several
{\it smoothing lengths}.  Gas properties for individual particles are
then estimated by summing over nearest neighbors, as in 
\begin{eqnarray}\label{eqn: sph}
\rho_i &=& \sum_j m_j W({\bf x}_i - {\bf x}_j , h_i, h_j), \nonumber \\
{d{\bf v}\over dt} = -{1\over\rho} \nabla P &\Longrightarrow &
   {d{\bf v}_i\over dt} = -\sum_j m_j \Biggl(2{{\sqrt{P_i P_j}\over{\rho_i
\rho_j}} + \Pi_{ij}}\Biggr) \nabla_i W,\;\;{\rm and} \nonumber \\
 {du\over dt} = -{P\over\rho} \nabla \cdot {\bf v} &\Longrightarrow &
   {du_i\over dt} = -{1\over\rho_i} \sum_j m_j \Biggl({{\sqrt{P_i P_j}\over
{\rho_i \rho_j}} + {1\over 2}\Pi_{ij}}\Biggr) ({\bf v}_i - {\bf v}_j)\cdot \nabla_i W,
\end{eqnarray}
where $h_i$ is the smoothing length, adjusted to keep a constant number 
of neighbors $N_{\it smooth}$ within $2h_i$, and $m_i$, ${\bf v}_i$,
$\rho_i$, $P_i$, and $u_i$ are the mass, velocity, density, pressure,
and specific thermal energy associated with each particle.
For the kernel function $W$, we take a spline function with compact support
within a radius of $2h_i$ given by (\cite{mon85})
\begin{equation}
W(r,h) \equiv {1\over \pi h^3}
\left\{
\begin{array}{ll}
1-1.5(r/h)^2+0.75(r/h)^3 & 0\leq r/h< 1, \\
0.25[1-(r/h)]^3 & 1\leq r/h< 2, \\
0 & r/h\geq 2,
\end{array}
\right.
\end{equation}
then take the average over particles $i$ and $j$ according to \cite{her89}
\begin{equation}
W({\bf x}_i - {\bf x}_j , h_i, h_j) = 
{1\over 2}[W({\bf x}_i - {\bf x}_j , h_i) + W({\bf x}_i - {\bf x}_j , h_j)].
\end{equation}
In equation (\ref{eqn: sph}), $\Pi_{ij}$ is the artificial viscosity given by
\begin{equation}
\Pi_{ij} = {q_i \over \rho_i^2} + {q_j \over \rho_j^2},
\end{equation}
with
\begin{equation}
q_i = 
\left\{
\begin{array}{ll}
\alpha h_i \rho_i c_i |\nabla \cdot {\bf v}|_i + \beta h_i^2 \rho_i |\nabla \cdot {\bf v}|_i^2 & {\rm for}\; |\nabla \cdot {\bf v}|_i < 0 \\
0 & {\rm for}\;|\nabla \cdot {\bf v}|_i \geq 0
\end{array}
\right.
\end{equation}
%$$\Pi_{ij} = {{-\alpha \mu_{ij} \bar{c}_{ij} + \beta \mu^2_{ij}}\over{\bar\rho_{ij}}}, $$
%$$\mu_{ij} = {({\bf v}_i - {\bf v}_j)\cdot\nabla({\bf x}_i - {\bf x}_j)\over
%{\bar{h}_{ij} (|{\bf x}_i - {\bf x}_j|^2/\bar{h}_{ij}^2 + \eta^2)}}$$
Additionally, we set $\Pi_{ij}=0$ for particle $i$ if ${\bf v}_i\cdot {\bf x}_i \geq 0$.
The parameters $\alpha\approx \beta\approx 0.5$ are
user-defined parameters which add dissipation in the presence of shocks.
The use of symmetrized quantities insures momentum conservation.
The version of SPH used here is quite similar to the one in \cite{her89} and 
\cite{kat96}.  Other choices are possible for various aspects of the SPH
calculations, in particular for the smoothing procedure and for
the artificial viscosity (see \eg \cite{her89} and \cite{mon92a}).

We incorporate an optional lower limit on the smoothing length,
specified as some fraction (typically $1/4$) 
of the gravitational softening length $\epsilon_{\rm grav}$
(\cite{evr88}).  This only has an effect on scales smaller than 
$\epsilon_{\rm grav}$ which are unresolved anyway, but prevents 
the use of very small timesteps which slows the calculation.

A single TreeSPH timestep, from $t^n$ to $t^{n+1}$, proceeds as follows:
\begin{itemize}
\item Begin with {\bf x}$^n$, {\bf v}$^n$, $u^n$ at $t^n$.
\item Advance {\bf x}$^n\;\rightarrow\;${\bf x}$^{n+1/2}$ using {\bf v}$^n$.
\item Compute {\bf a}$_{grav}^{n+1/2}$ using {\bf x}$^{n+1/2}$.
\item Estimate $\hat{\bf v}^{n+1/2}$ using {\bf a}$^{n-1/2}$
\item Determine $h^{n+1/2}$ using {\bf x}$^{n+1/2}$ by requiring $N_{\it smooth}$ neighbors.
\item Compute $\rho^{n+1/2}$ using {\bf x}$^{n+1/2}$ and $h^{n+1/2}$.
\item Advance $u^n\;\rightarrow\;u^{n+1/2}$ by a semi-implicit scheme (see \S\ref{sec: eth} below).
\item Compute {\bf a}$_{hydro}^{n+1/2}$ using {\bf x}$^{n+1/2}$, $h^{n+1/2}$, $u^{n+1/2}$ and $\hat{\bf v}^{n+1/2}$.
\item Advance {\bf v}$^n\;\rightarrow\;${\bf v}$^{n+1}$ using {\bf a}$_{total}^{n+1/2}$.
\item Advance {\bf x}$^{n+1/2}\;\rightarrow\;${\bf x}$^{n+1}$ using {\bf v}$^{n+1}$.
\item Advance $u^{n+1/2}\;\rightarrow\;u^{n+1}$ by a semi-implicit scheme (\S\ref{sec: eth}).
\end{itemize}

The use of time-centered quantities maintains errors at $O(\Delta t^2)$.
The parallelization of this algorithm is the focus of the remainder
of this section.

 \subsection{Parallelization Strategy}

A parallel implementation of SPH must be able to detect and compute the
contribution to SPH sums from neighboring particles $j$ which can be on 
different processors from particle $i$.
There are various strategies to handle this problem. 
The simplest method is to obtain information from another processor whenever it 
is required; however, the communication overhead is prohibitive because
a single processor must idly wait for the other processor to pass along the
information when it becomes ready, and there is usually significant overhead in
establishing interprocessor communication.  This problem can be alleviated by
using asynchronous communication, by which a given processor continues
to perform other work while waiting for information.
In PTreeSPH, we adopt an alternate strategy of using synchronous 
communications to build
a {\it locally essential particle list} (LEPL), a
collection of all particles
from other processors required to perform the local processor's SPH computations.
The second difficulty arises because the LEPL (obtained at the beginning of each
timestep) contains information updated
to the previous timestep, whereas SPH sums require current timestep information.
Thus we have developed a request-send algorithm to update gas particle
information during the timestep using the minimum possible 
amount of communication.
Communication benchmarks for our approach are comparable
to those of the asynchronous code presented in \cite{war94}.

To properly load-balance by the equal-work scheme, one must keep track of
the computational work done for each SPH particle.  This turns out
to be much more complicated than in the case of gravity alone where there
is a single computation (particle-cell interaction) which takes the
bulk of the time.
Thus we simply neglect this issue for now; that is, we use the
same particle list from the $N$-body domain decomposition, and
compute the pressure forces on the subset of those which are SPH particles.
It turns out that the SPH load balancing is still quite good 
(see \S\ref{sec: bench} and Table~\ref{table: bench}), which
is not surprising considering that denser regions will tend to be 
more computationally expensive for both gravity and SPH.

  \subsection{Building the Locally Essential Particle List}

The locally essential particle list (LEPL) is at the heart of our
parallel implementation of SPH.  By collecting a complete list of particles
required for a local SPH computation, we effectively
isolate the parallel portion of the code from the computation
of physical quantities, in a manner analogous to the locally
essential trees in the $N$-body portion.

In the case of building the locally essential trees, each processor
obtains the spatial extent of every other processor and therefore
is able to use the opening criterion to determine the tree 
information required by
every other processor.  For the LEPL, the situation is more complicated,
as the particle information required depends sensitively on the (often 
highly variable) smoothing region of particles near a given processor's boundary.
Obtaining all particles within the maximum extent of a
processor's smoothing region was found to be prohibitive in both
communication time and memory, requiring large numbers of unnecessary particles 
to be transmitted and stored.
Instead, we have developed an algorithm by which each processor
obtains only those particles required for the LEPL from the other
processors, and no (or very few) more.

The SPH formalism of \cite{her89} averages the gather and scatter kernels to 
insure symmetry of forces between every
pair of particles, thereby conserving momentum.  
Thus the algorithm must perform both a {\it gather search} to
obtain all non-local particles which fall within a local particle's
smoothing region, and a {\it scatter search} to find all the
non-local particles whose smoothing regions enclose local particles.
In all operations, only gas particles are considered, while
dark matter particles are ignored.  

\begin{figure}[hp]
\plotone{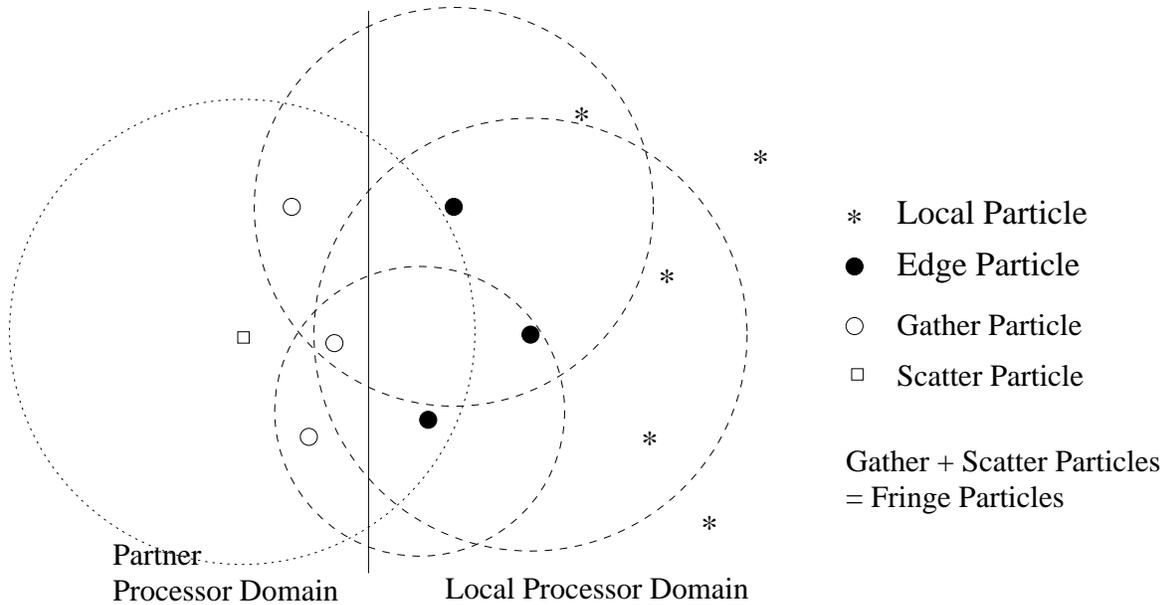}
\caption
{A locally essential particle list (LEPL).  Most local particles
(asterisks) require no information from other processors.
Edge Particles (filled circles), however, have smoothing volumes which
extend into the neighboring processor's domain.  A search of Edge
Particles' smoothing volumes identifies the Gather particles (open circles).
A scatter search must also be done to insure that the Scatter particles
(open square) are identified as well.  The Gather and Scatter particles
together form the Fringe particles, which are then sent back to
the local processor
for use in computations.  After all Fringe particles have been collected
from every other processor, SPH computations may proceed analogously to a
serial code, except when data updates are required.}
\label{fig: lepl}
\end{figure}

The LEPL building algorithm (see Figure~\ref{fig: lepl})
proceeds simultaneously on every processor, with communications
done synchronously with the partner processor when required.
\begin{itemize}
\item Identify particles whose search region extends into the partner 
processor's domain, which we call {\rm Edge particles}.
\item Exchange {\rm Edge particles} with partner processor.  Now the local
processor possesses a list of {\rm Partner Edge particles}.
\item Gather Search: For each {\rm Partner Edge particle}, search and tag 
{\rm Fringe particles}, \ie local neighbors of {\rm Partner Edge particles}.
\item Scatter Search:
\begin{itemize}
\item Tag all particles which are local {\rm Edge particles} but have not
already been tagged as {\rm Fringe particles}.
\item Find maximum smoothing length of those particles.
\item For each {\rm Partner Edge particle}, tag all particles within maximum
scatter search region as {\rm Fringe particles}.
\end{itemize}
\item Exchange tagged {\rm Fringe particles} with partner processor.
\item Append the received {\rm Fringe particles} to local particles to obtain the {\rm LEPL}.
\end{itemize}
A given processor performs this procedure with every other processor in
the simulation, although more distant processors will usually require
no exchange of particles.  This makes the communications scheme scale
as $N_P^2$, where $N_P$ is the number of processors, which is not as
optimal as the $N_P \log N_P$ scaling of the $N$-body parallelization.
Upon completion, a given processor possesses a complete set of
{\it Fringe particles}, which are
all particles outside the local domain which overlap in either a gather 
or scatter mode with one of the local particles.

The scatter search is somewhat inefficient because it uses a
maximum smoothing length, and often re-tags many of the particles
already identified as Fringe particles.  
Nevertheless it uses a relatively small portion of the LEPL computation time, so
we leave it for the future to improve this algorithm.

  \subsection{Data Updates}

In SPH, the value computed for the density is subsequently used to
compute $du/dt$ and advance $u$, which in turn is used to compute $d{\bf v}/dt$.
However, the LEPL contains only particle information updated to
the previous timestep.  Thus after each processor completes a
computation, a {\it data update} must be done to obtain
the current values of Fringe particle properties from other processors.
Unlike with locally essential trees, communication
must occur during the timestep as well as at the beginning.

To facilitate this, each gas particle carries not only an overall identification
number, but also a {\it local} identification number, which is
sequential on each processor.  When received from another processor, a
particle's local ID is encoded with the number of its host processor.
To update a physical quantity in 
a Fringe particle, we then use the following algorithm, again running
on all processors simultaneously between synchronous communications:
\begin{itemize}
\item Collect a list of local IDs of particles which are in the partner processor's domain.
\item Exchange these {\it request IDs} between processors.
\item Compile the required data for the list of request IDs.
\item Exchange the requested data.
\item Update the Fringe particles with the received data.
\end{itemize}
\rm

During each timestep, a data exchange is performed after any
computation which recomputes physical quantities (density, smoothing
length, etc.) for local particles.  In this manner each processor
always uses the most up-to-date information to compute smoothed
averages of SPH quantities.

  \subsection{Variable Smoothing Lengths}

Since smoothing lengths are varied to keep a fixed number of particles
within each particle's smoothing region, an algorithm must be used
to guarantee that a given particle is obtaining all Fringe particles within
its current smoothing region.  
Using the previous step's smoothing length will lead to 
errors, since the smoothing length may very well increase at the current 
timestep and thus all Fringe particles will not be properly identified in the 
gather search.

At the first timestep, all smoothing lengths are adjusted to
encompass $N_{\it smooth}$ particles {\it within the local
processor domain}.  For interior particles, this smoothing length 
is the true smoothing length; for Edge particles, this
smoothing length is guaranteed to be equal to or greater than
the true smoothing length, since it will find less particles
than actually exist within its search volume.
Thus the LEPL will contain more Fringe particles than necessary.
After building the LEPL, the smoothing length of Edge particles may be 
readjusted downwards to the correct value.

While this algorithm would work at all timesteps, the overcounting
of Fringe particles reduces efficiency.  To avoid this,
we perform the following algorithm 
at the end of each timestep: First, a data update
is performed to obtain all the Fringe particles' current positions.  
Then the smoothing
lengths of Edge particles are re-adjusted.  This prediction will
be much closer to the actual value since it includes Fringe particles
(unlike at the first timestep),
but is guaranteed to err on the side of undercounting the number of
particles within the search region, and thus overestimating the
smoothing length.  After the slightly overestimated smoothing length is used 
in collecting the LEPL, once again it can be readjusted downwards to
the proper value.

  \subsection{Grid-Tree Neighbor Searching}

In TreeSPH, neighbors are 
accumulated once per timestep and stored in memory, then recalled for each sum; 
despite this, the CPU time taken for neighbor finding is typically $\sim 1/3$ of the 
entire simulation time.  For parallel machines, storing a 
list of $\sim 30$ neighbors per particle is prohibitive,
since code performance degrades with increased memory
usage (see \S\ref{sec: memory}).
This necessitated the development of a significantly faster
algorithm using a hybrid grid-tree data structure for neighbor finding.
With grid-tree searching, neighbors can be found
whenever needed (typically on six separate occasions during a timestep), 
thereby eliminating the need for storing neighbor lists in memory.
Despite the repetition, the fraction of computation time required for
neighbor finding is comparable to that in TreeSPH.

\begin{figure}[hp]
\plotone{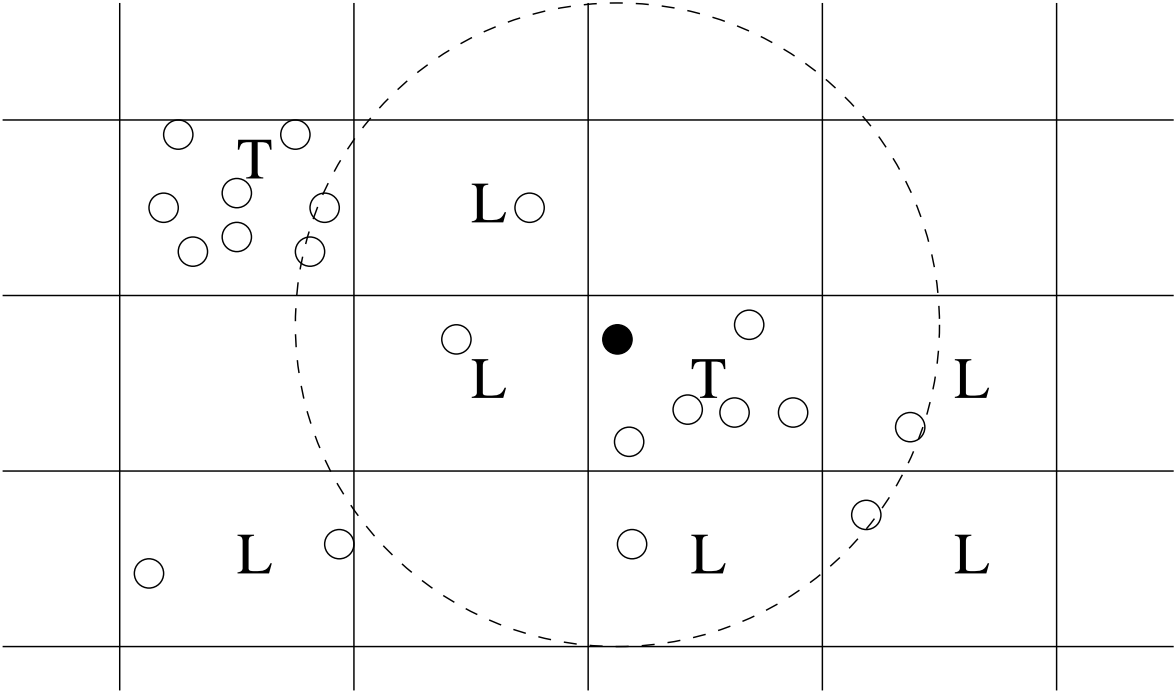}
\caption
{A schematic representation of the grid-tree data structure.  In high-density
grid cells (marked with a ``T") the particles within that cell are organized 
into a tree structure, while in lower-density grid cells (marked with a ``L")
the particles are left as a linked list.  When searching for neighbors,
the algorithm performs a treewalk in the dense grid cells, and cycles through
the linked list in less dense grid cells.  Thus the worst-case (highly 
clustered) scaling remains at $O(N\log N)$, while most of the volume utilizes 
the faster grid-linked list search.}
\label{fig: neighbor}
\end{figure}

The grid-tree scheme combines a grid-linked list search with a
tree search to obtain the speed advantages of both.  A grid-linked list 
search bins particles onto a grid, then makes a linked list of particles
within each grid cell.  Searching is then done by first identifying
which grid cells to search through, then cycling through each grid cell's
linked list and checking if each particle is within the search region.
The grid-linked list search is up to five times faster than the tree search method 
when all particles have roughly the same smoothing length.
However, since the search time scales as $O(N^2)$ where
$N$ is number in a densest grid cell, this method rapidly deteriorates
in clustered situations typical of astrophysical applications.  

A tree search walks down the tree by
determining if the particle's search region extends into a given
tree cell.  If so, the cell is ``opened" into its subcells, and
the test is repeated on the subcells, and so on until all neighbors are found.
TreeSPH used a pure tree search, which scales as $O(N\log N)$
regardless of clustering.

In the grid-tree structure in PTreeSPH, we replace the linked list 
with a tree structure when the number of particles in a grid cell
exceeds a threshhold value, typically $\sim 50$.  The searching
routines automatically detects whether a grid cell contains a tree
or linked list, and performs the appropriate search (see Figure~\ref{fig: neighbor}).
In unclustered regions we achieve the benefits of a grid-linked
list, while in clustered regions the search time still scales
as $O(N\log N)$.  For cosmologically clustered situations,
typical speedup is a factor of $\ga 2$.

By using the grid-tree scheme and writing the code in $C$ 
(generally a faster language than FORTRAN when handling data structures),
we have achieve relative speedups of factors of five over TreeSPH neighbor 
searching, removing the need to store neighbor lists.

  \subsection{Multiple Timestepping}

PTreeSPH uses individual timesteps for all particles, thereby
concentrating the computational effort in regions where higher resolution
is required.  For the dissipationless particles, we use the criterion
from \cite{kat96}, in which a particle's timestep is given by the
minimum of $\eta (\epsilon_{\rm grav}/v)$ and $\eta (\epsilon_{\rm grav}/a)^{1/2}$, where $v$ is the particle velocity, $a$ is the acceleration, and
$\eta$ is a user-specified number, typically $\sim 0.4$.

For the gas particles there is an additional timestep constraint 
set by the Courant condition (see \cite{her89}):
\begin{equation}
\Delta t_i = {\cal C} {h_i\over{h_i |\nabla\cdot{\bf v}_i| + c_i 
		+1.2\; (\alpha c_i + \beta h_i |\nabla\cdot{\bf v}_i|)}},
\end{equation}
where ${\cal C}\approx 0.3$ is the Courant number and $c_i$
is the sound speed.
Particles are binned according to
their timestep size into powers-of-two subdivisions of the largest timestep.
At any given small timestep each particle
falls into one of three categories:  (1) ``Active", meaning the given particle
needs a recalculation of its force in this small timestep,
(2) ``Inactive", meaning the given particle does NOT need recalculation of
its force,
and (3) ``Semi-active" (gas particles only), 
meaning the particle itself is inactive but
it has an SPH neighbor which is active.  

To maintain synchronous communication, all processors are assigned
the smallest timestep of the entire system.  Domain decomposition is
not redone, but the locally essential tree is rebuilt and the
gravitational force is calculated for all active particles.
To compute hydrodynamical forces, data updates are done to allow local 
advancing of 
the positions and velocities of Fringe particles.  All the
smoothing lengths and densities are recalculated, even for inactive particles.
Subsequent computations of $du/dt$ and the acceleration are
only performed on active and semi-active particles 
(required for their scatter contribution to active particles).  
For inactive particles, all physical quantities are interpolated 
to be current at each small timestep.

To maintain good load balancing between synchronous communication of these 
small timesteps, we tally the gravitational work of all active particles at
each small timestep and add that to the overall work for that processor.
This method is suboptimal since the tree exchange and tree 
construction overhead remain 
identical even though there may be only a small number of active particles, 
but it turns out the load balancing is still reasonable.
We leave it for the future to develop a partially asynchronous communication 
algorithm which may improve load balancing.

  \subsection{Memory Considerations}\label{sec: memory}

Memory considerations are important for the efficiency of a parallel code.
By placing the largest possible number of particles on each processor, the 
amount of time spent performing local computations increases relative
to the amount of time spent for computations on remote particles.
This means that a relatively smaller amount of remote information is
required, and hence overhead for communication time is reduced.

In PTreeSPH, the SPH calculation reuses the memory used to store the
$N$-body tree structure.  There are three major
uses for memory in the $N$-body treecode:  Storing the particle information,
storing the local tree, and storing the locally essential tree information from other 
processors.  The latter two are no longer needed once the gravitational
force has been computed, so the memory can be reused to store the
Fringe particles and grid-tree structure.  PTreeSPH has its own
memory handler routines which suballocate the memory out for tree cell 
storage or SPH storage as required.  Thus other than the increased memory
required to store the particles' hydrodynamical information (which roughly
doubles the amount of memory for particle information),
PTreeSPH actually requires no more memory than an $N$-body version
of the code.

On the T3D with 8 Megawords (64 Megabytes) per processor, simulations
may run with up to $\sim 64$~K particles per processor.  This is somewhat
smaller than $\sim 100$~K particles per processor possible with 
the pure $N$-body treecode, due to the increased memory requirements
for SPH information.
Currently PTreeSPH is roughly equally limited by memory and time allocation
constraints.  Newer architectures promise to scale both memory and performance
roughly equally, so we anticipate this will continue to be the case
in the near future.

  \subsection{Additions for Cosmological Simulations}\label{sec: eth}

PTreeSPH is currently set up to handle periodic boundary
conditions, radiative cooling, and heating from a parameterized
photoionizing background.
We have already described the periodic implementation of the
$N$-body portion of the code in \S \ref{sec: ewald}.
For SPH, the code must detect
neighbors on other processors accounting for the periodic nature of the simulation volume.
To do this, we first note that domain decomposition does not
cut across periodic boundaries, so it is only when Fringe particles are
required that periodicity becomes important.  Thus periodic boundaries affect
SPH only during the building of the LEPL.
We utilize the fact that a Fringe particle's position can be
different (in a periodic sense) from the original particle's position.
Whenever a Fringe particle position is received, either during the construction
of the LEPL or during a data update, the position of the particle is
adjusted to be closest possible to the local domain.  Subsequent neighbor
finding and SPH computations can then be done with no regard for 
periodic boundaries.  The only difficulty arises when a particle
appears multiply in a given processor's fringe, and such rare cases are detected
and Fringe particles are replicated when necessary; in practice this 
never occurs when the number of processors $N_P\geq 8$.

Radiative cooling and photoionization heating are implemented as described
in \cite{kat96} (\S 3).  Collisional excitation, collisional ionization, recombination, 
free-free emission and Compton cooling are all included, as well as heating 
provided by a user-input parameterized ionizing background.
Ionization equilibrium is assumed, but not thermal equilibrium.
PTreeSPH successfully reproduces the cooling and heating curves shown in 
\cite{kat96} figures~1 and~2; the reader is referred there for further details.

Since the cooling timescales can be considerably shorter than the timestep
as determined by the Courant condition, the thermal
energy cannot be integrated explicitly with the Courant timestep.
Instead, we use a Newton-Raphson method to implicitly solve
the following equation for $u^{n+1/2}$:
\begin{equation}
f(u^{n+1/2})\equiv u^{n+1/2}-u^n-{\Delta t\over 2}(\dot{u}^{n+1/2}+\dot{u}^n)=0
\end{equation}
where $\dot{u} = \dot{u}_{ad} + \dot{u}_{rad}$ is made up of adiabatic
and radiative contributions.  The values of $u^n$ and $\dot{u}^n$ are known
from the previous step, and $\dot{u}_{ad}^{n+1/2}$ is a stable enough function
of $u^{n+1/2}$ to be determined using a predictor-corrector method.
On the other hand, $\dot{u}_{rad}^{n+1/2}$ can be a rapidly varying 
function of $u^{n+1/2}$.  
Thus the adiabatic portion may be integrated explicitly, but the 
radiative portion must be integrated implicitly using the Newton-Raphson solver.
PTreeSPH's Newton-Raphson solver brackets the solution, then reduces the
acceptable range iteratively until the temperature is determined to
an accuracy of one part in $10^{-5}$.
In cases where a Newton-Raphson iteration would place the solution outside
the acceptable range, a bisection iteration is used instead;
the algorithm is adapted with extensive modifications
from Numerical Recipes (\cite{pre92}).
We have found this technique to be quite stable, and requires less computation
than integrating $\dot{u}_{rad}^{n+1/2}$ over very small timesteps given by
the cooling timescale.

Eventually we plan to add star formation (\cite{kat92a}; \cite{mih94}) with feedback.
Since each processor is
locally self-contained after the LEPL has been built, this addition
will be as straightforward as in the serial version of TreeSPH.

%%%%%%%%%%%%%%%%%%%%%%%%%%%%%%%%%%%%%%%%%%%%%%%%%%%%%%%%%%%%%%%%%%%%%%%%%%%%%%
\section{Testing PTreeSPH} 

  \subsection{Spherical Collapse}

The collapse of a spherically-symmetric cloud in three dimensions is a
good test of a hydrodynamical code's ability to handle problems with 
large dynamic range in space and time.
Here we compare the results of PTreeSPH to TreeSPH, which in turn 
agree quite well with one-dimensional finite element calculations
of this problem (\cite{her89}; \cite{evr88}).  We begin with a 
self-gravitating gas sphere of radius $R$ and total mass $M_T$, randomly
sampled by 4096 particles with a density profile
\begin{equation}
\rho = {M_T \over {2\pi R^2 r}}.
\end{equation}
The gas is initially isothermal with specific energy density $u=0.05 G M_T/R$,
and the ratio of specific heats is $\gamma = 5/3$.
The artificial viscosity parameters chosen were
$\alpha=\beta=0.5$.  The gravitational tolerance parameter was
taken to be $\theta=1.0$ for PTreeSPH, and $\theta=0.7$ for TreeSPH
(which give roughly identical gravitational force accuracies).  
The gravitational softening length is $0.1 R$, and smoothing lengths
were varied to keep $N_{\rm smooth}=32$ neighbors within a smoothing volume
(in the TreeSPH run, there was a tolerance of $\pm 3$ neighbors).
The Courant number was taken to be 0.3, and the largest system timestep
was taken to be $\Delta t_s = 0.022$ 
(in the adopted system units of $G=R=M_T=1$).  We do not include radiative
processes.

\begin{figure}[hp]
\plotone{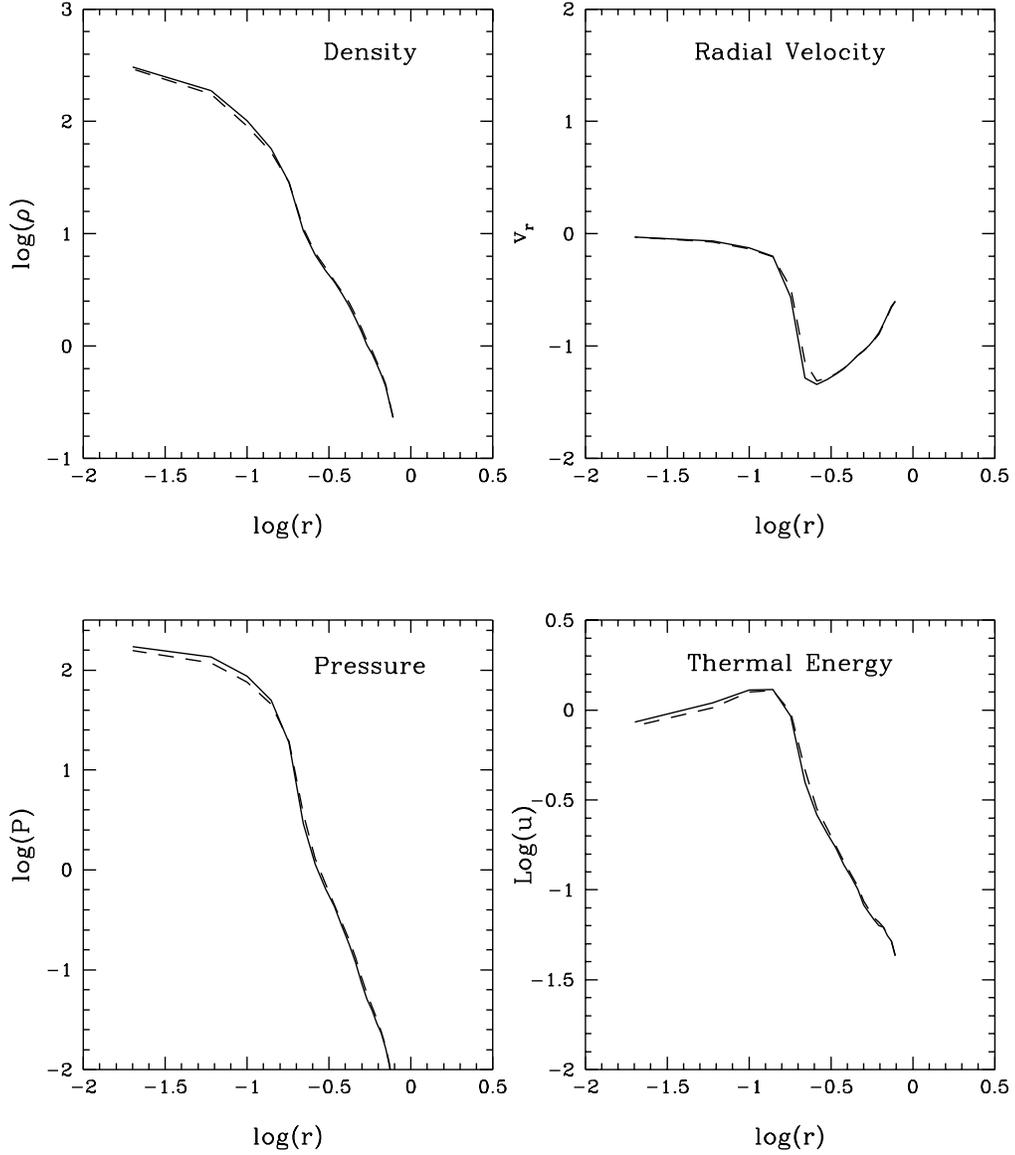}
\caption
{Spherical collapse of $\rho\propto 1/r$ sphere at $t=0.88$, where
the infall shock is most clearly visible.  The PTreeSPH results (solid line)
are quite similar with the TreeSPH results (dashed line),
which in turn agree with a 1-D finite difference calculation (see HK).
The shock is resolved over two to three smoothing lengths, with no discernible
postshock oscillations.}
\label{fig: coll}
\end{figure}

The resulting collapse produces a shock most prominently seen at roughly 
$t=0.88$.  Figure~\ref{fig: coll} shows the distribution of density,
thermal energy, pressure, and radial velocity as a function of radius.
The solid curve is the PTreeSPH
result, while the dashed curve is the TreeSPH result.
They are quite similar over a dynamic range in 
density of over 1000, and over a (maximum) dynamic range in time of 32.  
The shock is resolved over a few smoothing lengths, with no discernible
postshock oscillations.
Differences are attributable to variations in the gravity 
implementation and the use of a $\pm 3$ particle
tolerance on $N_{\rm smooth}$ in TreeSPH.
Results at other times are likewise very similar to TreeSPH results.

This PTreeSPH test was performed on four processors on an SGI Power Challenge.
Domain decomposition splits the problem directly through the densest
region of the simulation, where smoothing lengths are quite varied.
The agreement of the density profile of the two codes down to the
lowest radii is evidence that the Fringe particles are being
identified correctly (both gather and scatter particles) and all
required information is being exchanged appropriately.

  \subsection{Zel'dovich Pancake}

\begin{figure}[hp]
\plotone{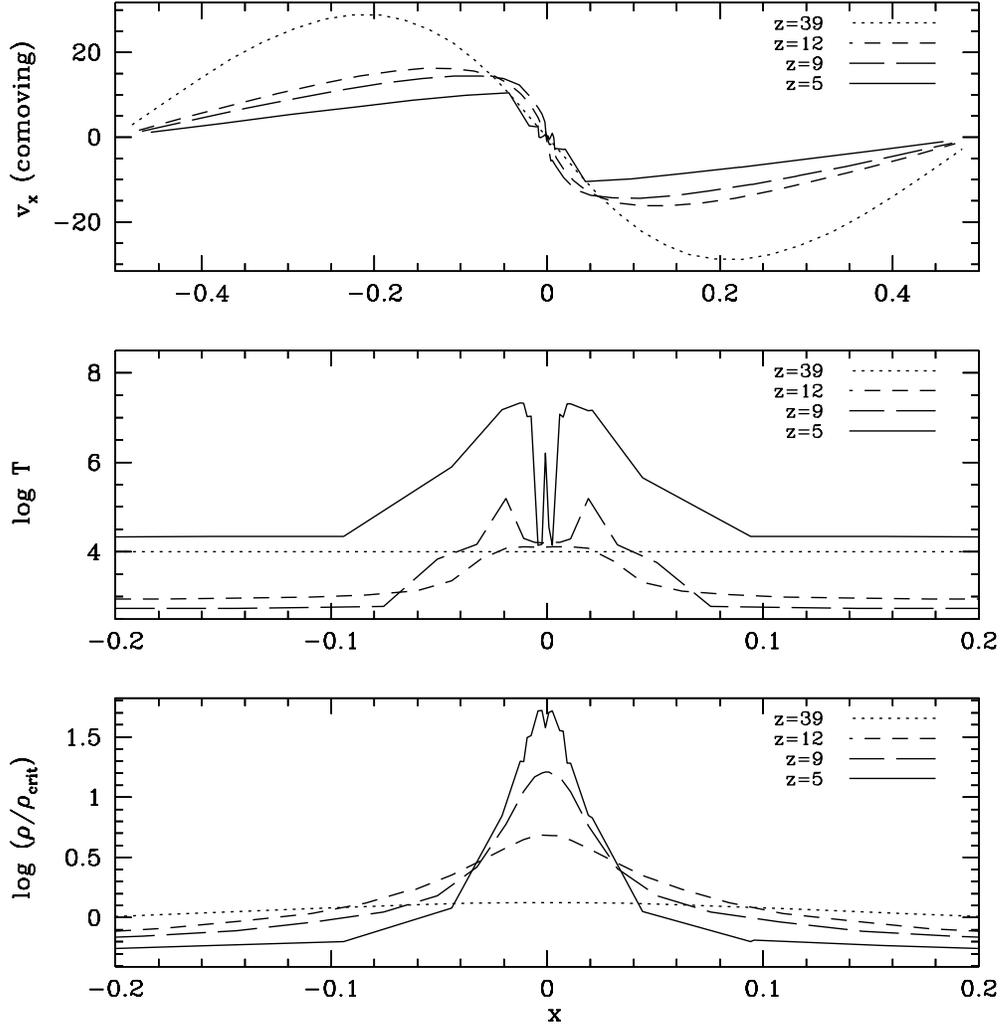}
\caption
{Zel'dovich pancake collapse of a sheet of gas showing the $x$-direction
velocity, temperature and density binned along the $x$-axis.
Dotted line shows initial state at $z=39$, a sinusoidal perturbation in 
displacement and velocity.
Short dashed line shows the state at $z=12$, as the initial shock has
formed.  Long dashed line shows the state at $z=9$,
showing the dual-shock front pattern with the central region maintained at
$10^{4.2}$~K by radiative cooling.  The final state at $z=5$ is shown
as the solid line, with multiple shocks due to secondary infall
and the background gas heated due to photoionization.}
\label{fig: zeld}
\end{figure}

The collapse of a Zel'dovich pancake (\cite{zel70}; \cite{bry94}) 
provides an interesting and relevant test 
for cosmological simulations.  The initial conditions are that of a 
sinusoidal perturbation in velocity and displacement from a lattice.
Figure~\ref{fig: zeld} (dotted line) shows $T$,
$v_x$, and $\rho$ versus $x$ in the initial state at $z=39$
(the temperature and density plots focus on the central region where
the interesting physics occurs at later times).
We use $32^3$ gas particles (hydrogen fraction by mass $X_H=0.76$)
in an $L=22.222$~Mpc (comoving) box with $H_0 = 50\;\kmsMpc$, starting at $z=39$ and ending at $z=5$.
We take $\epsilon_{\rm grav}=0.01 L$, $\theta=0.7$, and take large timesteps of 
$\Delta t=11.4\times 10^6$~years.  For
the hydrodynamics, we set $\gamma=5/3$, $N_{\rm smooth}=32$, ${\cal C}=0.3$, $\alpha=0.5$ and
$\beta=0.5$.  We also set a floor on the smoothing
length of $0.25 \epsilon_{\rm grav}$.  We include radiative cooling
and at late times turn on an ionizing background given by a power law with
\begin{equation}\label{eqn: Jnu}
J(\nu) = 10^{-22}(\nu_L/\nu)F(z)\;\;\Junits\;,
\end{equation}
with a redshift dependence 
\begin{equation}
F(z) = 
\left\{
\begin{array}{ll}
4/(1+z) & {\rm for}\;3<z<6, \\
1 & {\rm for}\;2<z<3.
\end{array}
\right.
\end{equation}
The simulation was performed on 8 processors of the Cray T3D at the 
Pittsburgh Supercomputing Center.

The resulting collapse initially follows linear theory, dominated by the 
gravitational force.  
The error in $v_x$ during this phase is $\lesssim 0.1$\%.  The gas cools
adiabatically from $10^4$~K due to cosmological expansion, with
slight variations due to differences in density.

After the caustic forms, hydrodynamical effects become dominant and a 
central shock region develops, as shown by the short dashed lines in 
Figure~\ref{fig: zeld} ($z=12$ at this time).  
Radiative processes efficiently cool the
outer shock envelope, so the temperature in the entire shock region is
maintained at $T \approx 10^{4.2}$~K.  As the shock strength builds,
the characteristic dual-shock pattern emerges, as shown be the state
of the system at $z=9$ in  Figure~\ref{fig: zeld} (long dashed line), while
the region between the shock fronts remains at $T \approx 10^{4.2}$~K
due to radiative cooling.
Figure~\ref{fig: zeld} (solid line) shows the final state at $z=5$,
where a more complicated structure has developed due to secondary
infall and resultant shock heating.
The background gas is heated by the ionizing background
to $T\approx 10^{4.5}$~K, where heating, radiative cooling, 
and adiabatic cooling due to cosmological expansion are in equilibrium.

While there are no analytical predictions for this problem in the nonlinear
regime, the general features are in good agreement with expectations in a
problem where shock heating, radiative cooling, and photoionization all
play major roles in governing the evolution of the system.

  \subsection{Cosmological Simulation: Comparison With TreeSPH}\label{sec: cosmo}

As a final test, we redid the TreeSPH simulation described in \cite{wei97}
using PTreeSPH.  This is a cosmological simulation of the Standard Cold
Dark Matter (CDM) model with $\Omega=1$, $H_0=50\;\kmsMpc$, $\Omega_b=0.05$,
and $\sigma_{\rm 16,mass}=0.7$, in a 22.222~Mpc (comoving) cube with
$32^3$ dark matter and $32^3$ gas particles.  The ionizing background
used is given in equation~(\ref{eqn: Jnu}).  The simulation was started at $z=49$
and run to $z=2$, using 762 large timesteps of 
$\Delta t = 3.3\times 10^6$~years, with small timesteps as low as $\Delta t/4$
(timesteps down to $\Delta t/16$ were allowed but never required).  
We took $\epsilon_{\rm grav}=20$~kpc and set a floor on the SPH smoothing length 
of 5~kpc.  Other SPH and cooling parameters were taken to be the same as in the
Zel'dovich pancake test.  We performed this run on 16 processors
of the Cray T3D at Pittsburgh Supercomputing Center.

\begin{figure}[hp]
\plotone{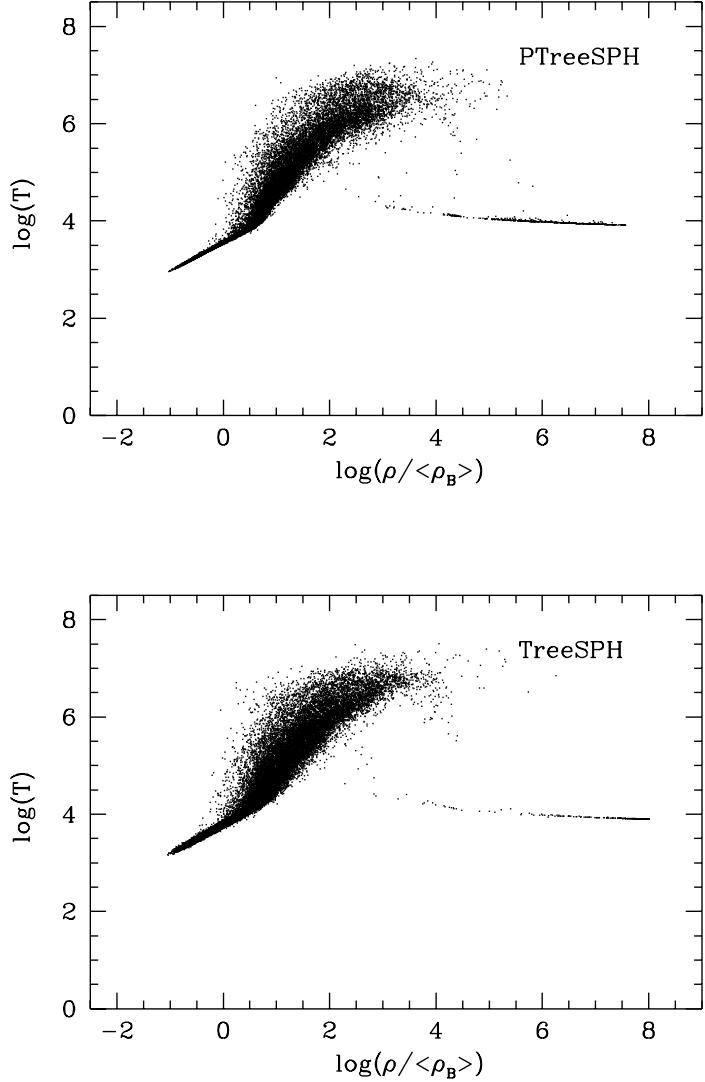}
\caption
{Scatter plot showing density $\rho$ vs. temperature $T$ for every SPH
particle in the $2\times 32^3$ run described in \ref{sec: cosmo}.  
Top panel shows PTreeSPH results, bottom panel
shows TreeSPH results.  Both codes show a reservoir of underdense
cool gas heated by photoionization, a reservoir of shock-heated gas
of moderate density, and a cool high-density component corresponding
to collapsed objects.  The last component shows some differences between
the two codes, with PTreeSPH having more particles in collapsed objects.
This arises from systematic overestimation of acceleration in the
linear regime with TreeSPH, causing more shock heating at early times
which delays the initial formation of small clumps. }
\label{fig: cosmo1}
\end{figure}

Figure~\ref{fig: cosmo1}a (top panel) shows a scatter plot of density (in units of the mean
baryonic density) vs. temperature for all gas particles in the simulation
at $z=2$.
The particles fall into three main categories, as described in \cite{kat96}:
Low-density gas cooled adiabatically by cosmic expansion; overdense,
shock-heated gas which generally surrounds large halos; and high-density
gas which has radiatively cooled to $T\sim 10^4$K and lies in dense clumps.
The last group is what we associate with galaxies.

Figure~\ref{fig: cosmo1}b shows the same scatter plot for the identical
TreeSPH run, reproduced from Figure~3 of \cite{wei97}.
While most general features are similar, there is a 
noticeable difference in the number of particles lying in cold, collapsed
clumps, particularly of intermediate density ($\rho \ga 10^3$).
This arises because TreeSPH systematically overestimates
accelerations in the linear phase as described in \S~\ref{sec: ewald},
hence it heats gas at an earlier time and prevents the early formation of
small collapsed objects.  This early heating was very pronounced in a comparison
of TreeSPH and PTreeSPH runs of the Zel'dovich pancake test described
above: TreeSPH produced an initial shock at $z\approx 15$, while
PTreeSPH produced an initial shock at $z\approx 11$.  Comparisons
of $\rho$ vs. $T$ plots for this cosmological simulation at various
redshifts also showed more hot particles in the TreeSPH run at early times.
PTreeSPH handles the linear regime more accurately by
including quadrupole Ewald terms and using a smaller tolerance parameter
at early times.  Despite the difference in dense particles, 
visualization showed that 
the major structures formed in the PTreeSPH and TreeSPH runs are all
very similar.

Another difference from Figure~\ref{fig: cosmo1}
is that the reservoir of low-density, photoionization-heated
gas shows less scatter in the PTreeSPH run than in the TreeSPH run.
The difference likely arises from the accuracy of the Newton-Raphson solvers
implemented in the two code.  
The temperature of a given SPH particle depends on the thermal energy
and the electron density, but the electron density in turn depends
on the temperature through the cooling and heating rates.  
Thus at every iteration of the Newton-Raphson solver, 
the mutual dependence of $T$ and $n_{\rm elec}$ 
must be iterated until a convergent solution is achieved.  
In TreeSPH, only one iteration
is used; however, this is sometimes insufficient to
achieve convergence, especially 
in the low-density regime where heating is
sensitively balanced by adiabatic cooling from cosmological expansion.
PTreeSPH iterates until convergence to obtain 
a tighter solution for the temperature in this regime.

%%%%%%%%%%%%%%%%%%%%%%%%%%%%%%%%%%%%%%%%%%%%%%%%%%%%%%%%%%%%%%%%%%%%%%%%%%%%%%
\section{Performance Characteristics of PTreeSPH on a Cray T3D}
  \subsection{Timing Results}\label{sec: bench}

For benchmarking purposes, we ran the cosmological simulation with $64^3$ 
dark matter and $64^3$ gas particles described in \cite{kat96} (similar
to the one in the previous section except with eight times as many particles),
beginning at $z=2$ and continuing for ten large timesteps.
At $z=2$, the simulation is already quite clustered and uses eight 
small timesteps per large timestep, so these benchmarks actually 
represent those for 80 steps.  We ran on eight processors of the Cray T3D
at the Pittsburgh Supercomputing Center.
This run is optimized to have the
most number of particles per processor (64K) that memory constraints will allow.

We divide the simulation into six
main routines which take up nearly all the computation time:

\begin{enumerate}
\item Gravity parallel overhead, which includes domain decomposition and locally essential tree building,
\item Gravity force calculation, which includes local tree building,
\item LEPL Building,
\item Data Updates,
\item Smoothing Length Adjustment, and
\item Hydrodynamical force calculation, which includes neighbor finding.
\end{enumerate}

\vskip 0.1in
\begin{tabular}{lcccc} \label{table: bench}
& Ave Time (sec)& \% Time & Load Balancing & Scalability \cr
Total & 3929.1 & 100.0 & 89.6 & 86.9 \cr
Gravity Parallelization & 15.3 & 0.4 & 99.5 & 62.6 \cr
Gravity Force & 1288.3 & 32.8 & 95.8 & 96.0 \cr
LEPL Building& 134.6 & 3.4 & 97.2 & 62.6 \cr
Data Updates & 261.5 & 6.7 & 64.5 & 62.6 \cr
$h_{\rm smooth}$ Adjustment & 423.7 & 10.8 & 85.4 & 88.9 \cr
SPH Force & 1333.9 & 33.9 & 88.7 & 88.9 \cr
\end{tabular}
\vskip 0.1in
\noindent
Table \ref{table: bench}: Benchmarks for $2\times 64^3$ run
averaged over ten large timesteps.  Load balancing is defined in 
equation~(\ref{eqn: load}), and scalability is described in \S\ref{sec: scale}.
\vskip 0.2in

Table~\ref{table: bench} shows the wallclock time 
taken per timestep per processor, averaged over ten large timesteps,
along with the corresponding fraction of the 
total computational time for each of the six routines.
The gravity and SPH computations each take about one-third of the total time,
with the rest divided among parallel overhead and smoothing length 
adjustment.  About 12\% of the total time is not accounted for, which 
represents I/O and initialization times.
The parallel overhead for the gravity computation is negligible,
but for SPH it takes $\sim 10$\% of the total time.  Much of this time goes
into determining which particles need to be exchanged, rather than 
interprocessor communication.  In addition, the
exchange of Fringe particle information must be done over every pair of
processors, making this is $N_P^2$ process (where $N_P$ is the number
of processors), rather than an
$N_P\log N_P$ process as for the gravity parallelization (see \cite{dub96}).
Despite these difficulties, the parallel overhead is still quite
reasonable, as roughly 90\% of the time is spent performing useful computations.

The average wallclock time taken per step with eight 
processors is around 3900~secs, or
a little over an hour per large timestep (\ie eight small timesteps).  
This translates to roughly 9 node-hours 
per large timestep for this $2\times 64^3$ simulation.  To evolve from $z=49$ to
$z=2$ (\ie the simulation done in \cite{kat96}) requires 762 large timesteps.
Since the early part
of the simulation runs faster (fewer small timesteps), we conservatively
estimate that this simulation could be run in 6000 node-hours of a Cray T3D.
A simulation with $10^7$ particles and half the resolution 
would conservatively require 32 times as much time, which
on a Cray T3E (approximately 4 times faster than a T3D)
translates to roughly 50,000 node-hours.
Alternatively, a $2\times 10^6$ particle simulation may be run to $z=0$
in around the same time.  Note that these benchmarks represent conservative
values based on incorporating cooling (which increases the SPH time by
up to a factor of two) and periodic boundary conditions (which increases
the overall code time by roughly a factor of two).  Simulations not
requiring these effects would be considerably faster, as would simulations
which require fewer small timesteps (\eg less clustered or lower resolution
simulations).

  \subsection{Load Balancing}

Ideally, a code should divide the computational work equally among all
processors.  However, because it is difficult to estimate the workload
{\it a priori}, some fraction of the time will be spent idly waiting for
the processor with the largest workload to complete its computations.
Such idle time occurs whenever an exchange of information is required, 
since all processors must be synchronized to perform exchanges in PTreeSPH's
synchronous communication scheme.  The six routines listed in the
previous section divide the computation into blocks between required exchanges.

We measure load balancing in each routine by computing the fractional 
amount of time spent idle while another processor performs computations.
More specifically, if $t_{\rm max}$ is the time taken by the processor
with the most work for a given routine, and $t_i$ is the time taken by
processor $i$, the load balancing for that routine is given by
\begin{equation}\label{eqn: load}
L = {1\over N_{\rm procs}} 
    \sum_{i=1}^{N_{\rm procs}} 1 - (t_{\rm max}-t_i)/t_{\rm max}
\end{equation}

Table~\ref{table: bench} shows the load balancing for each of
the six routines, averaged over the 10 timesteps.  The gravity
computation is load balanced at roughly a 96\% level, the 
SPH computation is around 89\%, and the smoothing length adjustment
is around 85\%.  The data update routine is the only one which does
not load balance well (64\%), but since the time in this routine is
fairly small, it does not significantly degrade the overall performance.
Overall, the code is load
balanced at roughly a 90\% level, meaning that only 10\% of the
wallclock time is spent with processors idle.
While these are not optimal values, they are quite
reasonable considering that the domain decomposition scheme is designed
to load balance the gravity portion only.  An asynchronous communication
scheme may improve load balancing, but the increased complexity of
the code may prove detrimental to the overall performance.  We
leave this for future investigations.

  \subsection{Scalability}\label{sec: scale}

\begin{figure}[hp]
\plotone{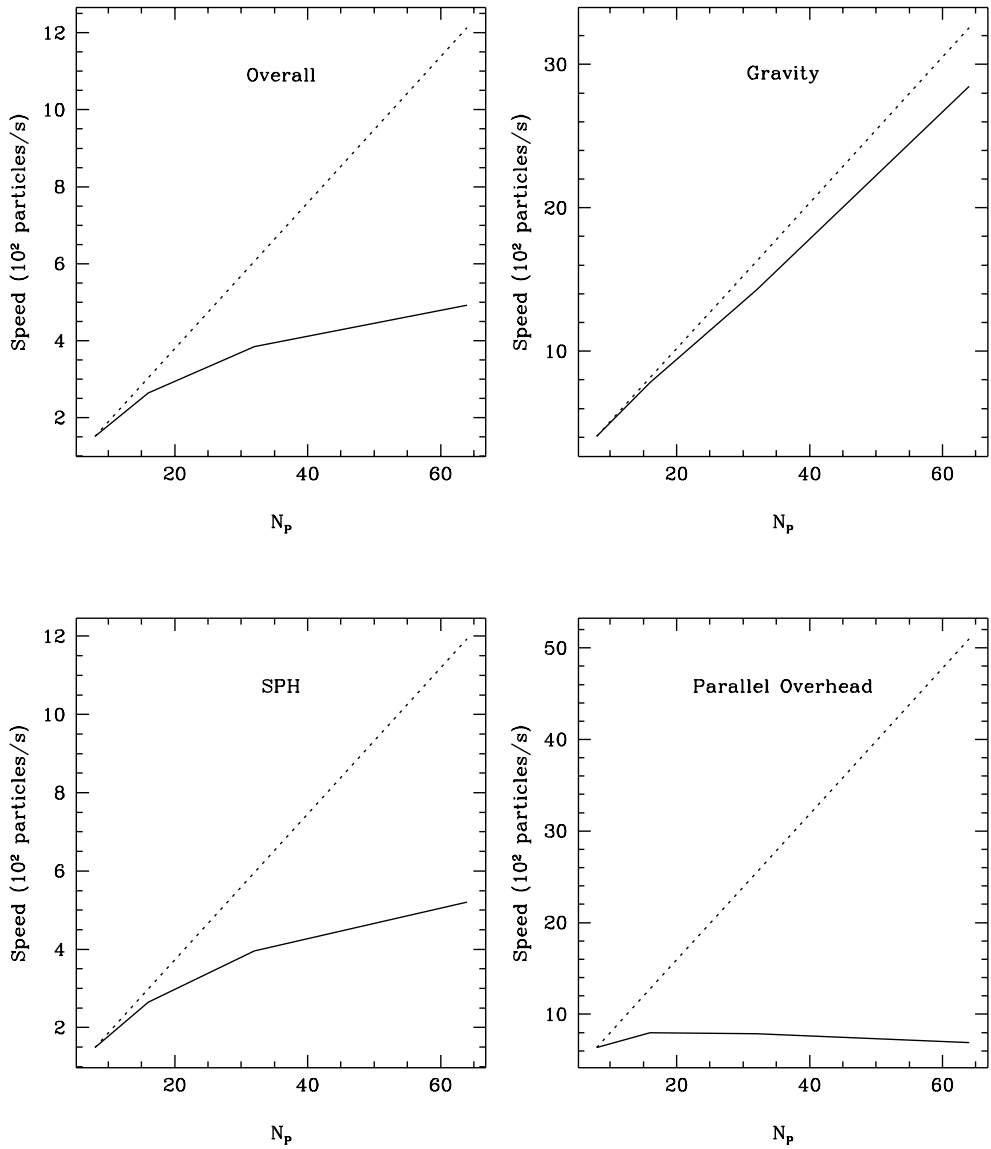}
\caption
{Scalability of various routines of PTreeSPH with number of processors,
for the run described in \S\ref{sec: scale}.  Solid lines represent
PTreeSPH timings at 8, 16, 32 and 64 processors.  Dotted lines show
ideal scaling from 8 processors.
Upper left shows the overall code scaling.  Upper right shows the
gravity portion scaling, which is quite good.  Lower left shows
scaling of the SPH portion, including smoothing length adjustment.
This is fairly poor up to 64 processors, but reasonable up to 16 processors.
Lower right shows the scaling of the parallelization routines, the
domain decomposition, locally essential tree building,
LEPL building and data updates.  This speed remains as a fairly constant
overhead for any number of processors.  The SPH and parallelization scaling
illustrates the efficiency gained by minimizing the number of processors
used for a given simulation.}
\label{fig: scale}
\end{figure}

Ideally, the speed of each routine should scale linearly with the number
of processors.  However, due to irregular domain sizes and increasing
sommunication overhead, the speed degrades with increasing numbers of
processors.  To test this, we consider 10-timestep runs as described above with
8, 16, 32 and 64 processors.  Figure~\ref{fig: scale} shows the scaling
with number of processors for three major components of the code:
the gravity portion, the SPH portion (which here includes smoothing length
adjustment), and the LEPL portion (which here includes data updates).
The last column of Table~\ref{table: bench} shows the scaling as a percentage
of ideal between 8 and 16 processors.

The gravity portion scales very well all the way from 8 to 64 processors,
suffering just a few percent loss of speed at each increment.  The
SPH portion does not scale nearly as well out to 64 processors;
however, between 8 and 16 processors, the loss is only $\sim 11$\% from ideal.  
The parallelization overhead stays roughly constant.
Overall, PTreeSPH's scaling is dominated by the SPH portion,
but between 8 and 16 processors the loss of speed is only $\sim 14$\%.
The scaling of ultimate interest is that between eight and one
processors, which we cannot directly measure, but we take the scaling
between 16 and 8 processors as a conservative estimate.
These graphs illustrate the importance of minimizing the number
of processors used for a given simulation.  
As memory constraints are alleviated in future machines, the scalability 
of PTreeSPH should improve.

\section{Summary and Future Work}

A parallel TreeSPH code for performing cosmological hydrodynamical
simulations has been presented.  The physics is incorporated in a manner
quite similar to the TreeSPH code (\cite{her89}; \cite{kat96}), but the algorithm
has been redesigned to handle difficulties unique to
the massively parallel environment.
The code is capable of performing cosmological simulations in a periodic
volume, including the effects of radiative cooling and heating from
a parameterized ionizing background.
The use of massively parallel supercomputers currently enables one to perform
simulations with roughly an order of magnitude more
particles than is practically possible on a vector supercomputer.
Moreover, with the current rate of advance in chip technology, the massively
parallel approach promises to provide rapid further increases in computational
power and therefore simulation size.  Currently, a cosmological hydrodynamical 
simulation of $10^7$ particles down to $z=2$
is feasible within a typical allocation of supercomputer time; alternatively,
one can explore models with many smaller simulations.  
Significantly larger simulations are possible for applications which do
not require periodic boundaries, or whose matter is less clustered
than in a cosmological simulation.

PTreeSPH has been implemented with primary concern for portability
and expandability, and secondary concern for optimal efficiency.
Nevertheless, the code achieves good load balancing ($\sim 90$\%) and
has fairly low communication overhead ($\sim 10$\%).
Compile-time flags have been implemented to easily include or exclude
various physical processes such as hydrodynamics, cooling, and cosmology.
The use of the MPI message passing
software allows for easy portability between parallel platforms; already
the code has been run successfully on the Cray T3D, the IBM SP2, and an SGI
Power Challenge.
The building of locally essential problems, both for gravity
and hydrodynamic forces, allows for straightforward incorporation of other
physical processes as desired.
Together with its fully Lagrangian nature, this will make PTreeSPH 
useful for a wide variety of astrophysical applications, including
cosmology, galaxy interactions and cluster formation.

%%%%%%%%%%%%%%%%%%%%%%%%%%%%%%%%%%%%%%%%%%%%%%%%%%%%%%%%%%%%%%%%%
\section{Acknowledgements}
We are grateful to Neal Katz, David Weinberg and Guohong Xu for
helpful comments and suggestions.
We acknowledge grants of computer resources by Pittsburgh
Supercomputing Center and the Cornell Theory Center.  This
work was supported by the Grand Challenge Cosmology Consortium,
NSF grant ASC 93-18185, and the Presidential Faculty Fellows Program.
%%%%%%%%%%%%%%%%%%%%%%%%%%%%%%%%%%%%%%%%%%%%%%%%%%%%%%%%%%%%%%%%%

\end{document}